\documentclass[11pt,a4paper]{article}

\usepackage{fullpage}
\usepackage{amsmath}
\usepackage{amssymb}
\usepackage{amsthm}
\usepackage{mathrsfs}
\usepackage{graphicx}
\usepackage{psfrag}
\usepackage{citesort}

\newcommand {\be} {\begin{eqnarray*}}
\newcommand {\ee} {\end{eqnarray*}}
\newcommand {\bea} {\begin{eqnarray}}
\newcommand {\eea} {\end{eqnarray}}

\newcommand{\bm}[1]{\boldsymbol{#1}}

\newcommand{\ave}[1]{\langle {#1} \rangle}

\newcommand{\pdiff}[2]{\frac{\partial{#1}}{\partial{#2}}}

\newcommand{\qedd}{\hfill $\blacksquare$}

\title{Barbero-Immirzi parameter, manifold invariants and Euclidean path integrals}
\author{\textbf{Tom$\acute{\mbox{a}}\check{\mbox{s}}$ Liko}\footnote{Electronic mail: tliko@math.ualberta.ca}\\
\\{\small \it Department of Mathematical and Statistical Sciences}\\
{\small \it University of Alberta}\\
{\small \it Edmonton, AB T6G 2G1, Canada}}

\begin{document}

\maketitle





\begin{abstract}

The Barbero-Immirzi parameter $\gamma$ appears in the \emph{real} connection formulation of gravity in terms of
the Ashtekar variables, and gives rise to a one-parameter quantization ambiguity in Loop Quantum Gravity.  In this
paper we investigate the conditions under which $\gamma$ will have physical effects in Euclidean Quantum Gravity.
This is done by constructing a well-defined Euclidean path integral for the Holst action with non-zero cosmological
constant on a manifold with boundary.  We find that two general conditions must be satisfied by the spacetime manifold
in order for the Holst action and its surface integral to be non-zero: (i) the metric has to be non-diagonalizable;
(ii) the Pontryagin number of the manifold has to be non-zero.  The latter is a strong topological condition, and
rules out many of the known solutions to the Einstein field equations.  This result leads us to evaluate the on-shell
first-order Holst action and corresponding Euclidean partition function on the Taub-NUT-ADS solution.  We find that
$\gamma$ shows up as a finite rotation of the on-shell partition function which corresponds to shifts in the energy
and entropy of the NUT charge.  In an appendix we also evaluate the Holst action on the Taub-NUT and Taub-bolt solutions
in flat spacetime and find that in that case as well $\gamma$ shows up in the energy and entropy of the NUT and bolt
charges.  We also present an example whereby the Euler characteristic of the manifold has a non-trivial effect on
black-hole mergers.

\end{abstract}

\hspace{0.35cm}{\small \textbf{PACS}: 04.20.Cv;04.20.Fy;04.70.Dy}


\section{Introduction}

Since its discovery within the context of canonical gravity \cite{barbero}, the Barbero-Immirzi parameter,
denoted $\gamma$, has remained an elusive one-parameter quantization ambiguity in loop quantum gravity
\cite{immirzi,rovthi}.  For details, see \cite{rovelli,ashlew,thiemann}.  This parameter can be fixed by
comparing the quantum theory to the semi-classical theory.  This procedure for fixing $\gamma$ has been done
by matching the Bekenstein-Hawking entropy of black holes with spacetime topology
$\mathbb{R}^{2}\times\mathbb{C}^{2}$, with $\mathbb{C}^{2}$ a compact two-manifold, to the corresponding isolated
horizon quantum geometry \cite{abck,abk,aepv,kbd}.  It turns out that $\gamma$ has the same value regardless of
the topology of the black hole.

The Einstein field equations admit more solutions in four dimensions than just the black holes whose event
horizons have topology $\mathbb{R}\times\mathbb{C}^{2}$; in particular, the NUT-charged spacetimes 
\cite{ntu,misner} have topologies that cannot be foliated by a time function and contain Misner
strings with a non-zero entropy \cite{hawhun}.  A comparison of the semi-classical and quantum geometry
descriptions of NUT-charged spacetimes would be of interest in order to provide an independent determination
of $\gamma$ that can be compared to the previous results for black holes \cite{abck,abk,aepv,kbd}.

Motivated by this interest, we want to know what effects, if any, $\gamma$ may have on the Euclidean path
integral.  In this paper we study the Euclidean Holst action \cite{holst} with non-zero cosmological constant,
and derive the semi-classical energy and entropy of Taub-NUT-ADS spacetime in the presence of $\gamma$.  This
can only be done in first-order formalism.  That said, we need to make sure that the first-order Holst-ADS
action satisfies two important conditions in order that we may be able to evaluate the on-shell Euclidean
Holst-ADS action: (I) \emph{The action has to have a well-posed action principle on a manifold with boundary};
and (II) \emph{the action has to be finite}.

Condition I leads us to consider the Dirichlet boundary value problem for a \emph{generic} first-order
action that includes curvature and torsion as functionals of the connection and coframe, and we present
a general prescription for determining the corresponding surface terms for which the first variation of the
action vanishes identically.  We then present, in an example, a derivation of the surface terms that are
necessary for a generalized Hilbert-Palatini action in four dimensions to be functionally differentiable; the
form of this functional is motivated by requiring consistency with the Hamiltonian theory coupled to fermions
when $\gamma=i$.  A special case of this action is the Holst action with non-zero cosmological constant -- the
main focus of study in this paper.  Some general properties of the Holst surface term are discussed in Section 3.
Condition II is also addressed in Section 3.  The ADS and Holst-ADS actions are finite without the need of adding
infinite counter-terms to the boundary integral, as is normally done in the infinite subtraction method
\cite{hawking1,gibhaw,balkra,lau,mann,manmar}.

Once the two conditions are shown to be satisfied by the Holst-ADS action, we are able to evaluate the Euclidean
Holst action on solutions to the field equations, and obtain the corresponding on-shell partition functions.  We
find that generically $\gamma$ shows up in the Euclidean path integral as a finite rotation of the on-shell partition
function, and this rotation corresponds to shifts in the energy and entropy of the spacetime.  This property of the
Holst-ADS action is a result of a strong topological condition that we find: the Holst term and its surface term are
non-zero if the Pontryagin number of the corresonding spacetime manifold is non-zero.  We confirm these results by
explicitly evaluating the Holst-ADS action and partition function on the Taub-NUT-ADS solution; $\gamma$ shows up
as a finite shift in the energy and entropy of the NUT charge.

We also include two appendices.  In Appendix 1 we evaluate the Euclidean Holst action on Taub-NUT spacetimes with
zero cosmological constant.  We show that $\gamma$ shows up as a shift in the energy and entropy of the NUT and bolt
charges, just as for the Taub-NUT-ADS solution.  In Appendix 2 we present an example which shows how a topological
invariant of the manifold can have non-trivial physical effects in gravity.  In particular, it is shown that the
presence of the Euler characteristic of the manifold can lead to violations of the second law during a black-hole
merging process.

\section{First-order action principle: generalities}

In the first-order formulation of general relativity (see e.g. \cite{ashlew}), the configuration space $\mathscr{C}$
is the pair $\{e,A\}$, consisting of the coframe $e$ and a connection $A$ valued in $SO(D)$ or $SO(D-1,1)$ depending
on the signature of spacetime.  The coframe determines the spacetime metric
$g_{ab}=\eta_{IJ}e_{a}^{\phantom{a}I} \otimes e_{b}^{\phantom{a}J}$ and spacetime volume form $\bm{\epsilon}=e^{0} \wedge \cdots
e^{D-1}$, where $\epsilon_{I_{1}\cdots I_{D}}$ is the totally antisymmetric Levi-Civita tensor.  In this paper, spacetime
indices $a,b,\ldots\in\{0,\ldots,D-1\}$ are raised and lowered using the metric $g_{ab}$ and internal indices
$I,J,\ldots\in\{0,\ldots,D-1\}$ are raised and lowered using the flat metric $\eta_{IJ}$.  The connection determines
the curvature two-form $\Omega_{\phantom{a}J}^{I}=dA_{\phantom{a}J}^{I}+A_{\phantom{a}K}^{I} \wedge A_{\phantom{a}J}^{K}=\frac{1}{2}
R_{\phantom{a}JKL}^{I}e^{K} \wedge e^{L}$, with $R_{\phantom{a}JKL}^{I}$ the Riemann tensor.  The Lagrangian density is denoted
$\mathcal{L}$; this is a \emph{functional}, by which we mean a map from the space of functions $(e,A)$ to $\mathbb{R}$.
The functional derivative of $\mathcal{L}$ with respect to a function $\varphi=\varphi(x^{a})$ is denoted
$\bm{\Upsilon}_{\varphi}\equiv \delta\mathcal{L}/\delta\varphi$.  The internal Hodge dual is denoted by $\star$.  In this
paper, we write differential forms without indices.

\subsection{Functional differentiability of the generic first-order action}

Let us begin with the following.

\noindent{\bf Proposition 1.}
\emph{Let $\Omega=dA+A\wedge A$ and $T=de+A\wedge e$ be (resp.) the curvature and torsion
of a $D$-dimensional manifold $\mathcal{M}$ with boundary $\partial\mathcal{M}$, with $A$
the connection and $e$ the coframe.  Let $\mathcal{L}[e,A,\Omega,T]$ be the Lagrangian density,
a $D$-form in spacetime.  The first-order action for an arbitrary diffeomorphism invariant theory
of pure gravity on the configuration space $\mathscr{C}=\{e,A\}$,}
\bea
I = \frac{1}{16\pi}\int_{\mathcal{M}}\mathcal{L}[e,A,\Omega,T]
              - \frac{(-1)^{D}}{16\pi}\oint_{\partial\mathcal{M}}\bm{\Upsilon}_{\Omega} \wedge A
              + \bm{\Upsilon}_{T} \wedge e \, ,
\label{foaction1}
\eea
\emph{is functionally differentiable.}

\noindent{\bf Proof.}
Take the variation of $\mathcal{L}$:
\be
\delta\mathcal{L}
&=& \bm{\Upsilon}_{e} \wedge \delta e
  +  \bm{\Upsilon}_{A} \wedge \delta A
  + \bm{\Upsilon}_{\Omega} \wedge \delta\Omega
  + \bm{\Upsilon}_{T} \wedge \delta T\\
&=& \bm{\Upsilon}_{e} \wedge \delta e
    + \bm{\Upsilon}_{A} \wedge \delta A
    + \bm{\Upsilon}_{\Omega} \wedge (d\delta A + 2A \wedge \delta A)
    + \bm{\Upsilon}_{T} \wedge (d\delta e - e \wedge \delta A + A \wedge \delta e)\\
&=& (-1)^{D}d(\bm{\Upsilon}_{\Omega} \wedge \delta A) + (-1)^{D}d(\bm{\Upsilon}_{T} \wedge \delta e)
    + \left[\bm{\Upsilon}_{e} + \bm{\Upsilon}_{T} \wedge A - (-1)^{D}d\bm{\Upsilon}_{T}\right] \wedge \delta e\\
& & + \left[\bm{\Upsilon}_{A} + 2\bm{\Upsilon}_{\Omega} \wedge A - (-1)^{D}d\bm{\Upsilon}_{\Omega}
    - \bm{\Upsilon}_{T} \wedge e\right] \wedge \delta A \, .
\ee
The action will be functionally differentiable if the total derivative is cancelled, and the equations of motion
\bea
\bm{\Upsilon}_{e} + \bm{\Upsilon}_{T} \wedge A - (-1)^{D}d\bm{\Upsilon}_{T} = 0\\
\bm{\Upsilon}_{T} \wedge e + (-1)^{D}d\bm{\Upsilon}_{\Omega} - 2\bm{\Upsilon}_{\Omega} \wedge A - \bm{\Upsilon}_{A} = 0
\label{eqsmotion}
\eea
are satisfied.  From the \emph{Fundamental Theorem of Exterior Calculus}, the boundary term
\bea
I_{\partial\mathcal{M}} = \frac{(-1)^{D}}{16\pi}\oint_{\partial\mathcal{M}}\bm{\Upsilon}_{\Omega}\wedge A + \bm{\Upsilon}_{T} \wedge e
\eea
follows.  This completes the proof.  \qedd

From the identity $d^{2}=0$, terms involving higher derivatives in the configuration variables $e$ and $A$
do not show up in the action (\ref{foaction1}).   The action (\ref{foaction1}) is therefore generically
first-order in derivatives of $e$ and $A$.

\subsection{Example: Generalized Hilbert-Palatini action in four dimensions}

If we take the Wilsonian point of view on effective field theories, then we need to add all possible $D$-forms
to the action for gravity that are diffeomorphism invariant.  In four dimensions, this implies that in addition
to the Hilbert-Palatini and cosmological terms, we should include in the action the Holst term
$\theta_{1}\smallint e\wedge e\wedge\Omega$, as well as the characteristic classes: the Euler class
$\theta_{2}\smallint\star\Omega\wedge\Omega$, the Pontryagin class $\theta_{3}\smallint\Omega\wedge\Omega$, and
the Nieh-Yan class $\theta_{4}\smallint T\wedge T-e\wedge e\wedge\Omega$.  See e.g. \cite{fmt,rezper}.  Here,
$\theta_{1}=1/\gamma$ with $\gamma$ a non-zero real constant -- the so-called Barbero-Immirzi parameter
\cite{barbero}.

It was pointed out in \cite{mercuri1,mercuri2}, however, that in the case when fermions are coupled to
gravity, the Holst term needs to be replaced with the Nieh-Yan term.  This extension of the Holst term
is necessary in the presence of non-zero torsion so that the corresponding Hamiltonian with arbitrary
$\gamma$ reduces to the Ashtekar-Romano-Tate Hamiltonian \cite{art} when $\gamma$ is set equal to the
imaginary unit.  Therefore we consider here the Nieh-Yan term \emph{in place of} the Holst term in our
action principle (with $\theta_{4}=1/\gamma$).  When torsion is zero the Nieh-Yan term reduces to the
Holst term.  This makes the topological origin of the Holst term in the first-order action manifest!

From Proposition 1, then, we have the following:

\noindent{\bf Corollary 2.}
\emph{The generalized Hilbert-Palatini action for general relativity with cosmological constant
$\Lambda=3\varepsilon/\ell^{2}\in\mathbb{R}$ ($\varepsilon\in\{-1,1\}$), Barbero-Immirzi parameter
$\gamma\in\mathbb{R}\setminus0$ and $\theta_{2},\theta_{3}\in\mathbb{R}$ on a four-dimensional manifold
$(\mathcal{M},e,A)$ with boundary $\partial\mathcal{M}$,}
\bea
I &=& \frac{1}{16\pi}\int_{\mathcal{M}}\star(e \wedge e) \wedge \Omega - 2\Lambda\bm{\epsilon}
    + \frac{1}{\gamma}(T \wedge T - e \wedge e \wedge \Omega) + \theta_{2}\star\Omega \wedge \Omega
    + \theta_{3}\Omega \wedge \Omega\nonumber\\
& & - \frac{1}{16\pi}\oint_{\partial\mathcal{M}}\star(e \wedge e) \wedge A - \frac{1}{\gamma}e \wedge de
    + 2\theta_{2}\star\Omega \wedge A + \theta_{3}\Omega \wedge A \, ,
\label{foaction3}
\eea
\emph{is functionally differentiable.}

\noindent{\bf Proof.}
For the Nieh-Yan-Holst term, we find that $\bm{\Upsilon}_{T}=T$ and $\bm{\Upsilon}_{\Omega}=-e\wedge e$
so that the surface term has density $T\wedge e-e\wedge e\wedge A=de\wedge e+A\wedge e\wedge e-A\wedge
e\wedge e=de\wedge e$.  For the Euler term we find that $\bm{\Upsilon}_{\Omega}=2\star\Omega$ so that
the surface term has density $2\star\Omega\wedge A$.  For the Pontryagin term we find that
$\bm{\Upsilon}_{\Omega}=\Omega$ so that the surface term has density $\Omega\wedge A$.  The surface term
in the action (\ref{foaction3}) follows. \qedd

The action (\ref{foaction2}), with $\theta_{1}=\theta_{2}=\theta_{3}=\theta_{4}=0$, was previously studied
within the context of black-hole mechanics.  In these works, the spacetimes under consideration include
four-dimensional Einstein-Maxwell theory with zero cosmological constant \cite{afk,abl}, non-trivial
matter couplings \cite{ashcor,corsud,acs}, higher-dimensional flat and ADS spacetimes \cite{klp,apv},
Gauss-Bonnet gravity \cite{likboo1} and supergravity with $p$ form matter couplings
\cite{likboo2,boolik,liko1}.

In this formalism, the topology of the boundary in the action principle is taken to be
$\partial\mathcal{M}\cong M_{1}\cup M_{2}\cup\Delta\cup\mathcal{I}$, with $\Delta$ a $(D-1)$-dimensional
null hypersurface equipped with a null normal and a degenerate metric with signature $(0+\ldots+)$.
$M_{1}$ and $M_{2}$ are partial Cauchy surfaces that extend from $\Delta$ to the boundary $\mathcal{I}$
at infinity; $M_{1}$ and $M_{2}$ intersect $\Delta$ and $\mathcal{I}$ in $(D-2)$-surfaces $\mathbb{C}^{D-2}$.
Adding the surface term $\oint_{\mathcal{I}}\star(e \wedge e)\wedge A$ to the action at $\mathcal{I}$ and
fixing the geometry of $\Delta$ are sufficient conditions for the action to be functionally differentiable
and for the zeroth and first laws of black-hole mechanics to be satisfied.  In particular, \emph{all}
the conserved charges are defined locally at the horizon $\Delta$; these include the non-monopolar
(dipole) charge of the five-dimensional black ring solution \cite{liko1,cophor}.

Subsequently it was shown that the action is finite on asymptotically flat spacetimes \cite{aes,ashslo},
and that a partition function can be given for Euclidean metrics \cite{likslo} without having to add infinite
counter-terms to the boundary \cite{likslo}.  It was then shown that the the Holst term and its surface term
together are finite \cite{corwil}.  In this paper we further explore properties of the Holst action and
corresponding Euclidean path integral on a manifold with boundary, in the presence of a non-zero cosmological
constant.

\section{Holst action with cosmological constant}

In this paper we will focus particular attention to ADS spacetimes with $\Lambda<0$, but the mathematical
results obtained also apply to de Sitter spacetimes with $\Lambda>0$.  Therefore we consider the Holst
action in the presence of a \emph{non-zero} cosmological constant
$\Lambda=3\varepsilon\ell^{-2}\in\mathbb{R}\setminus0$, with $\varepsilon\in\{-1,1\}$ and $\ell$ is
the de Sitter radius.  Specifically, we consider the action (\ref{foaction3}) with $T=0$ and with
$\theta_{2}=\theta_{3}=0$.

The first-order Holst action with cosmological constant $\Lambda=3\varepsilon\ell^{-2}\in\mathbb{R}\setminus0$
($\varepsilon\in\{-1,1\}$) and Barbero-Immirzi parameter $\gamma\in\mathbb{R}\setminus0$ on a four-dimensional
manifold $(\mathcal{M},e,A)$ with boundary $\partial\mathcal{M}$ is:
\bea
I = \frac{1}{16\pi}\int_{\mathcal{M}}\left[\left(\star - \frac{1}{\gamma}\right)(e \wedge e)\right] \wedge \Omega
    + \frac{6\varepsilon}{\ell^{2}}\bm{\epsilon}
    - \frac{1}{16\pi}\oint_{\partial\mathcal{M}}\star(e \wedge e) \wedge A - \frac{1}{\gamma}e \wedge de \; .
\label{foaction2}
\eea
From Corollary 2, this action is functionally differentiable.  The surface term here is the same as the one
that was previously found for the Holst action in flat spacetime \cite{corwil}.  This is because when
$T=de+A\wedge e=0$ we have $e\wedge de=e\wedge e\wedge A$.

An important property of the Holst surface term $e\wedge de$ is that it is identically zero for spacetimes
with metrics that can be put into \emph{diagonal} form.  To see this, consider the tetrad in components:
$e^{I}=e_{a}^{\phantom{a}I}dx^{a}$.  Differentiating the tetrad gives
$de^{I}=(\partial e_{a}^{\phantom{a}I}/\partial x^{b})dx^{b}\wedge dx^{a}$.  By direct substitution we get
\bea
e \wedge de = \left[e_{0}\left(\pdiff{e_{a}^{\phantom{a}0}}{x^{b}}\right) + e_{1}\left(\pdiff{e_{a}^{\phantom{a}1}}{x^{b}}\right)
+ e_{2}\left(\pdiff{e_{a}^{\phantom{a}2}}{x^{b}}\right) + e_{3}\left(\pdiff{e_{a}^{\phantom{a}3}}{x^{b}}\right)\right]
\wedge dx^{b} \wedge dx^{a}
\eea
If the metric is diagonal, then
\bea
e \wedge de
&=& e_{0}\left(\pdiff{e_{0}^{\phantom{a}0}}{x^{b}}\right) \wedge dx^{b} \wedge dx^{0}
    + e_{1}\left(\pdiff{e_{1}^{\phantom{a}1}}{x^{b}}\right) \wedge dx^{b} \wedge dx^{1}\nonumber\\
& & + e_{2}\left(\pdiff{e_{2}^{\phantom{a}2}}{x^{b}}\right) \wedge dx^{b} \wedge dx^{2}
    + e_{3}\left(\pdiff{e_{3}^{\phantom{a}3}}{x^{b}}\right) \wedge dx^{b} \wedge dx^{3} = 0 \; .
\eea
Of particular interest in general relativity are spacetimes with a `$t$-$\phi$' component in
their corresponding metrics.  For a general metric with $x^{0}$-$x^{3}$ cross term, $e\wedge de$
can be written in component form such that
\bea
e \wedge de = \left\lbrace e_{00}\left(\pdiff{e_{3}^{\phantom{a}0}}{x^{b}}\right)
              - e_{03}\left(\pdiff{e_{0}^{\phantom{a}0}}{x^{b}}\right)\right\rbrace dx^{0} \wedge dx^{b} \wedge dx^{3} \; .
\eea
In spherical coordinates $x^{a}\in\{\tau,r,\theta,\phi\}$, on a constant-$r$ hypersurface, the boundary term is:
\bea
e \wedge de = \left\lbrace e_{00}\left(\pdiff{e_{3}^{\phantom{a}0}}{\theta}\right)
              - e_{03}\left(\pdiff{e_{0}^{\phantom{a}0}}{\theta}\right)\right\rbrace d\tau \wedge d\theta \wedge d\phi \; .
\label{offdiagonal}
\eea
This expression can be used to evaluate the Holst surface term on solutions with a $t$-$\phi$ component in the metric.

For spacetimes with non-zero cosmological constant, another condition can be found on the Holst surface term.
Let us first substitute the equation of motion $de+A\wedge e=0$ into the action (\ref{foaction2}) to eliminate
the tetrad derivative in the surface term:
\bea
I = \frac{1}{16\pi}\int_{\mathcal{M}}\left[\left(\star - \frac{1}{\gamma}\right)(e \wedge e)\right] \wedge \Omega
    + \frac{6\varepsilon}{\ell^{2}}\bm{\epsilon}
    - \frac{1}{16\pi}\oint_{\partial\mathcal{M}}\star(e \wedge e) \wedge A - \frac{1}{\gamma}e \wedge e \wedge A \; .\nonumber\\
\label{foaction4}
\eea
Then with the equation of motion $e\wedge e=(\varepsilon\ell^{2}/6)\Omega$ the boundary term becomes (with
$\stackrel{!}{=}$ denoting \emph{equality on-shell})
\bea
-\frac{1}{16\pi}\oint_{\partial\mathcal{M}}\star(e \wedge e) \wedge A - \frac{1}{\gamma}e \wedge e \wedge A
\stackrel{!}{=} -\frac{\varepsilon\ell^{2}}{6\cdot16\pi}\oint_{\partial\mathcal{M}}\star \Omega \wedge A - \frac{1}{\gamma}\Omega \wedge A \; .
\eea
From the Fundamental Theorem of Exterior Calculus, this boundary integral can be written as a bulk integral:
\bea
-\frac{\varepsilon\ell^{2}}{6\cdot16\pi}\oint_{\partial\mathcal{M}}\star \Omega \wedge A - \frac{1}{\gamma}\Omega \wedge A
\stackrel{!}{=} -\frac{\varepsilon\ell^{2}}{6\cdot16\pi}\int_{\mathcal{M}}\star \Omega \wedge \Omega - \frac{1}{\gamma}\Omega \wedge \Omega \; .
\eea
Putting this in (\ref{foaction4}), we see that the Holst action can be written as a \emph{bulk} integral:
\bea
I = \frac{1}{16\pi}\int_{\mathcal{M}}\left[\left(\star - \frac{1}{\gamma}\right)(e \wedge e)\right] \wedge \Omega
    + \frac{6\varepsilon}{\ell^{2}}\bm{\epsilon} - \frac{\varepsilon\ell^{2}}{6}\star \Omega \wedge \Omega
    + \frac{\varepsilon\ell^{2}}{6\gamma}\Omega \wedge \Omega \; .
\label{finiteaction}
\eea
Written this way, the surface terms appearing in the action (\ref{foaction2}) are invariants of the manifold $\mathcal{M}$.
It follows from this form of the action that \emph{the Holst surface term is identically zero for manifolds that have zero
Pontryagin number}.  In addition, the Holst term itself can be written purely in terms of the connection by using the equation
of motion $e\wedge e=(\varepsilon\ell^{2}/6)\Omega$ to eliminate the tetrad.  We find that:
\bea
\frac{1}{\gamma}\int_{\mathcal{M}}e \wedge e \wedge \Omega
\stackrel{!}{=} \frac{\varepsilon\ell^{2}}{6\gamma}\int_{\mathcal{M}}\Omega \wedge \Omega \; .
\eea
Therefore we see that on-shell the Holst term will be identically zero if the Pontryagin number of the manifold is zero.

The first-order Holst action with negative cosmological constant is finite.  To see this, consider first the Einstein-Hilbert
and cosmological terms only.  Then, action (\ref{finiteaction}) with $\varepsilon=-1$ is precisely the action that was shown
to be finite for ADS spacetimes \cite{acotz}.  It was also shown in \cite{corwil} that the Holst term and its surface term
\emph{together} are finite.  Therefore, the action (\ref{finiteaction}) is finite.

Let us briefly summarize this section.  The Holst-ADS action is functionally differentiable and finite.  In order for the
Holst term and its surface term, and therefore $\gamma$ to be present in the on-shell action, the corresponding solution
has to have a non-diagonalizable metric and a non-zero Pontryagin number.  These are very restrictive conditions and rule
out many of the known spacetimes that are critical points of the action.  The Taub-NUT spacetime is known to have Pontryagin
number $2$ \cite{hawking2}.  Therefore, in Section 4 and in Appendix 1, we evaluate the Euclidean on-shell Holst-ADS and
Holst actions and partition functions (resp.) on the Taub-NUT-ADS spacetime with $\Lambda<0$ and the Taub-NUT and Taub-bolt
spacetimes with $\Lambda=0$.

\section{Euclidean path integrals and Taub-NUT-ADS spacetime}

\subsection{Partition functions and thermodynamics}

Let us consider the formal path integral
\bea
\mathcal{Z} = \int\mathcal{D}[\Psi]\mbox{exp}\left\lbrace -I[\Psi]\right\rbrace \, ,
\label{epath1}
\eea
with $I[\Psi]$ the Euclidean action and $\Psi$ a generic field variable.  Here, the measure
$\mathcal{D}[\Psi]$ includes all fields and not just the classical fields $\widetilde{\Psi}$
that satisfy the equations of motion $\delta I[\widetilde{\Psi}]=0$.  However, if the dominant
contributions to the partition function come from fields that are close to the classical fields,
then the action can be expanded in a Taylor series:
\bea
I[\widetilde{\Psi}+\delta\Psi] = I[\widetilde{\Psi}] + \delta I[\widetilde{\Psi},\delta\Psi]
                                 + \delta^{2}I[\widetilde{\Psi},\delta\Psi] + \ldots \; .
\eea
In order for the path integral $\mathcal{Z}$ to make sense, at least to second order in the Taylor series, we require
that the first term $I[\widetilde{\Psi}]$ be finite and that the linear term $\delta I$ vanish identically.
If these conditions are satisfied, then the \emph{on-shell} partition function is approximated by
\bea
\widetilde{\mathcal{Z}} = \mbox{exp}\left\lbrace -I[\widetilde{\Psi}]\right\rbrace \, ;
\label{epath2}
\eea
the average energy $\ave{E}$ and entropy $S$ are then given by
\bea
\ave{E} = -\pdiff{\ln\widetilde{\mathcal{Z}}}{\beta}
\quad
\mbox{and}
\quad
S = \beta\ave{E} + \ln\widetilde{\mathcal{Z}} \; .
\label{eands}
\eea
The physical meaning of the energy may differ based on the boundary conditions that are used, i.e. holding
the pressure or volume constant.

From Section 3, we know that the Holst-ADS action (\ref{foaction2}) is functionally differentiable and finite.
Therefore we may consider the Holst-ADS partition function
\bea
\widetilde{\mathcal{Z}} = \mbox{exp}\left\{-\frac{1}{16\pi}\int_{\mathcal{M}}\left[\left(\star
                          - \frac{1}{\gamma}\right)(e \wedge e)\right] \wedge \Omega
                          + \frac{6\varepsilon}{\ell^{2}}\bm{\epsilon}
                          - \frac{\varepsilon\ell^{2}}{6}\star \Omega \wedge \Omega
                          + \frac{\varepsilon\ell^{2}}{6\gamma}\Omega \wedge \Omega\right\}
\label{partitionfn}
\eea
for spacetimes with negative cosmological constant and non-zero Pontryagin number.  Let us therefore proceed
by evaluating the partition function (\ref{partitionfn}) on the Taub-NUT-ADS spacetime.

\subsection{Taub-NUT-ADS spacetime}

Here we consider the Taub-NUT-ADS spacetime.  The metric for four-dimensional \emph{Euclidean} Taub-NUT
spacetime, with negative cosmological constant $\Lambda=-3\ell^{-2}$, has line element
\bea
ds^{2} &=& V(r)\left[d\tau + 2N\cos\theta d\phi\right]^{2} + \frac{dr^{2}}{V(r)}
          + (r^{2} - N^{2})(d\theta^{2} + \sin^{2}\theta d\phi^{2}) \, ,\nonumber\\
V(r) &=& \frac{r^{2}-2Mr+N^{2}+(r^{4}-6N^{2}r^{2}-3N^{4})\ell^{-2}}{r^{2}-N^{2}} \, ,
\label{etaubnut1}
\eea
with $N$ the NUT parameter.  Regularity of the metric requires that the Euclidean time $\tau$ have
a period $\beta=8\pi N$.

A suitable tetrad of coframes for this spacetime is given by
\bea
e^{0} &=& \sqrt{V}d\tau + 2\sqrt{V}N\cos\theta d\phi \, ,
\quad
e^{1} = \frac{1}{\sqrt{V}}dr \, ,
\quad
e^{2} = \sqrt{r^{2} - N^{2}}d\theta \, ,\nonumber\\
e^{3} &=& \sqrt{r^{2} - N^{2}}\sin\theta d\phi \; .
\label{tnuttetrad1}
\eea
The Euclidean action for the Taub-NUT-ADS spacetime is then given by
\bea
I = I_{0} + \frac{64\pi^{2}N^{2}}{\gamma}\left(1 - \frac{2N^{2}}{\ell^{2}}\right) \, ,
\eea
with $I_{0}$ the on-shell action of the Taub-NUT-ADS solution without the Holst term (i.e. contributions
from the Einstein-Hilbert and cosmological terms only) \cite{mann,cejm}.  Substituting this in (\ref{epath1})
then gives the on-shell partition function
\bea
\widetilde{\mathcal{Z}} = \mbox{exp}\left\lbrace-I_{0} - \frac{64\pi^{2}N^{2}}{\gamma}\left(1 - \frac{2N^{2}}{\ell^{2}}\right)\right\rbrace \; .
\label{parfun1}
\eea
Whence the thermodynamic quantities are given by
\bea
\ave{E} = \ave{E}_{0} + \frac{N(\ell^{2}-4N^{2})}{\gamma\ell^{2}}
\quad
\mbox{and}
\quad
S = S_{0} + \frac{4\pi N^{2}}{\gamma}\left(1 - \frac{N^{2}}{\ell^{2}}\right) \, ,
\eea
with $\ave{E}_{0}$ and $S_{0}$ denoting (resp.) the average energy and entropy of the Taub-NUT-ADS solution
without the Holst term \cite{mann,cejm}.

We conclude that $\gamma$ appears in the thermodynamics of the Taub-NUT-ADS spacetime as a shift in the energy
and entropy of the NUT charge (provided that $\gamma$ is finite and real).  The quantities derived here are in
agreement with previous results found by Mann \cite{mann} and Chamblin \emph{et al} \cite{cejm} but with a finite
shift in the energy and entropy of the Taub-NUT-ADS spacetime; these shifts vanish in the limit when $\gamma$ is
taken to infinity.

\section{Summary and discussion}

Let us briefly summarize the main results that are presented in this paper.  We studied the properties of the
first-order Holst action with non-zero cosmological constant.  In particular, we showed that the spacetimes for
which $\gamma$ will be present in the on-shell action (\ref{finiteaction}), and hence in the partition function
(\ref{partitionfn}), are the ones that have non-diagonalizable metrics and non-zero Pontryagin number.  This led
us to evaluate (\ref{partitionfn}) on the Taub-NUT-ADS spacetime.  It was found that $\gamma$ shifts the energy
and entropy of the NUT charge.  The analogous results for the case where $\Lambda=0$ are presented in Appendix
1.  Some results regarding the Euler characteristic and black-hole mechanics are presented in Appendix 2.

The results found in this paper agree with recent results found by Durka and Kowalski-Glikman \cite{durkow,durka}.
Durka and Kowalski-Glikman derived the Noether-Wald charges for solutions of a constrained $BF$ theory with
$SO(3,2)$ symmetry, first introduced by Freidel and Starodubtsev \cite{fresta}.  They found that $\gamma$ shows
up in the Noether-Wald charges of solutions that have $\partial_{\theta}g_{t\phi}\neq0$.  For the Taub-NUT-ADS
solution, the energy and entropy are shifted by the same factor as found here, but with $1/\gamma\rightarrow\gamma$.
This is because, in the action that they studied, the Holst and Nieh-Yan terms appear seperately with different
weight factors: the Holst term has coefficient $1/\gamma$ while the Nieh-Yan term has coefficient
$(\gamma^{2}+1)/\gamma$, so their action is fundamentally different.  In particular, the limit $\gamma\rightarrow\infty$
cannot be taken in this action to recover the Einstein-Hilbert action.  Apart from this minor difference, the two
approaches are equivalent: $\gamma$ contributes to the on-shell partition function and to the Noether-Wald charges
through the Pontryagin number of the spacetime.

In this paper we considered initially the Holst action with a generic cosmological constant, and in particular
focused on the case when $\Lambda<0$.  However, in light of cosmological data, it would be of interest to also
study in detail the Holst action with $\Lambda>0$.  The expression (\ref{partitionfn}) is mathematically valid
for \emph{any} sign $\varepsilon=\mbox{sign}(\Lambda)$ of the cosmological constant.  An important step toward
determining how $\gamma$ affects e.g. the NUT-charged DS spacetimes \cite{cgm1,cgm2}, is to first prove the
finiteness of the Holst-DS action with $\varepsilon=1$.  Then (\ref{partitionfn}) will be a well-defined
partition function for DS spacetimes with non-vanishing Pontryagin number.  This approach may also reveal
new insights regarding the Kodama wavefunction with arbitrary real values of $\gamma$ \cite{randono1,randono2}.

In this paper we looked at the torsion-free case.  This field will be non-zero in the presence of fermion
couplings, and therefore should be included in the action.  Torsion-squared Lagrangian densities in the
first-order action have recently been studied in \cite{dtv,baeheh}; these terms are all consistent with
Corollary 2.  Note that in the presence of fermion couplings, $\gamma$ will appear in the chiral anomaly
\cite{chazan}.  Of particular interest would be to extend the formalism here to supergravity.  To study the
effects of $\gamma$ in supergravity, the supersymmetric Holst actions found by Kaul \cite{kaul} have to be
generalized to a manifold with boundary.  In practice, however, finding supersymmetric boundary terms
\emph{without imposing any boundary conditions on the fields} is difficult.  See \cite{belvan1,belvan2,belvan3}
for details.  Ideally we would like to have an action principle for supergravity that is invariant under the
\emph{off-shell} supersymmetry albegra; this suggests that we extend the supergravity action to a manifold
with boundary in superspace; such an action without boundary has been recently found by Gates Jr. \emph{et al}
\cite{gky} where it was found that $\gamma$ shows up in superspace as the complex component of the gravitational
constant.

It would also be of interest to determine the effects of $\gamma$ in quantum gravity by studying more general
Euclidean path integrals.  Because any topology may occur in classical and quantum gravity, one may have in
general a partition function that sums over all possible inequivalent topologies \cite{isham,harwit,carlip}.
In Section 3 we found that in order for $\gamma$ to be present in the partition function, the Pontryagin
number of the manifolds have to be non-zero, so only those manifolds will contribute to the formal sum.  In
the case of supergravity with fermion couplings, the existence of a spinor structure on $\mathcal{M}$ requires
that the second Stiefel-Whitney class of the manifold be non-zero \cite{ishpop}.  This condition places further
restrictions on the formal sum.

\section*{Acknowledgements}

The author wishes to thank Remigiusz Durka and Jerzy Kowalski-Glikman for very useful correspondence,
and for commenting on a draft of the manuscript.  The author also thanks Abhay Ashtekar and David Sloan
for discussions during the early stages of this project.  This work was supported by NSERC.  While at
Penn State University, the author was also supported in part by NSF grant PHY0854743, The George A.
and Margaret M. Downsbrough Endowment and the Eberly research funds of Penn State.

\appendix

\section{Taub-NUT and Taub-bolt spacetimes}

Here we consider the Taub-NUT spacetime.  The metric for four-dimensional \emph{Euclidean} Taub-NUT
spacetime, with zero cosmological constant, has line element
\bea
ds^{2} &=& V(r)\left[d\tau + 2N\cos\theta d\phi\right]^{2} + \frac{dr^{2}}{V(r)}
          + (r^{2} - N^{2})(d\theta^{2} + \sin^{2}\theta d\phi^{2}) \, ,\nonumber\\
V(r) &=& \frac{r^{2}-2Mr+N^{2}}{r^{2}-N^{2}} \, ,
\label{etaubnut2}
\eea
with $N$ the NUT parameter.  Regularity of the metric requires that $\tau$ have a period $\beta=8\pi N$.

A suitable tetrad of co-frames for this spacetime is given by
\bea
e^{0} &=& \sqrt{V}d\tau + 2\sqrt{V}N\cos\theta d\phi \, ,
\quad
e^{1} = \frac{1}{\sqrt{V}}dr \, ,
\quad
e^{2} = \sqrt{r^{2} - N^{2}}d\theta \, ,\nonumber\\
e^{3} &=& \sqrt{r^{2} - N^{2}}\sin\theta d\phi \; .
\label{tnuttetrad2}
\eea
Using (\ref{offdiagonal}), we find that the Euclidean action for the Taub-NUT spacetime
is given by
\bea
I = 4\pi MN - \frac{2\pi N^{2}}{\gamma} \; .
\eea
Substituting this in (\ref{epath2}) then gives the on-shell partition function
\bea
\widetilde{\mathcal{Z}} = \mbox{exp}\left(-4\pi MN + \frac{2\pi N^{2}}{\gamma}\right) \; .
\label{parfun2}
\eea
The thermodynamic quantities can now be calculated.  In particular, $M=N$ for the NUT charge and
substituting this into (\ref{parfun2}) gives the average energy and entropy
\bea
\ave{E} = N\left(1 - \frac{1}{2\gamma}\right)
\quad
\mbox{and}
\quad
S = 4\pi N^{2}\left(1 - \frac{1}{2\gamma}\right) \, ,
\eea
while $M=5N/4$ for the bolt charge and substituting this into (\ref{parfun2}) gives the average energy
and entropy
\bea
\ave{E} = \frac{5N}{4}\left(1 - \frac{2}{5\gamma}\right)
\quad
\mbox{and}
\quad
S = 5\pi N^{2}\left(1 - \frac{2}{5\gamma}\right) \; .
\eea
Therefore, $\gamma$ appears in the thermodynamics of Taub-NUT and Taub-bolt solutions with zero cosmological
constant as shifts in the energies and entropies of the NUT and bolt charges, just as we found for the NUT
charge in the Taub-NUT-ADS solution.

\section{Euler characteristic and the second law}

In this Appendix, let us present an example that illustrates how a topological invariant of the spacetime
manifold can have non-trivial physical effects on black-hole thermodynamics.  In particular, we will derive
an upper bound on $\theta_{2}$ that must be satisfied in order for the second law to hold when two black holes
merge.

The bound presented here is general and holds for solutions to the Gauss-Bonnet field equations in arbitrary
dimensions.  We will consider four-dimensional asymptotically flat black holes as a special case; in this case
the black holes must satisfy certain theorems and these can be used to put a tight upper bound on $\theta_{2}$
for the area theorem to hold when two Schwarzschild black holes merge \cite{liko2}.

For black holes of Gauss-Bonnet gravity with generic cosmological constant $\Lambda$, the first law of
black-hole mechanics holds with an entropy that is given by \cite{likboo1,jkm,crs}
\bea
\mathscr{S} = \frac{1}{4\pi}\oint_{\mathbb{C}}\mathbf{\tilde{\epsilon}}(1 + 2\theta_{2}\mathcal{R}) \, ;
\label{entropy4}
\eea
here $\mathcal{R}$ is the Ricci scalar of the horizon cross section $\mathbb{C}$ and $\mathbf{\tilde{\epsilon}}$
is the area $(D-2)$-form on $\mathbb{C}$.

Let us consider the merging of two black holes -- one with mass $m_{1}$ and the other with mass $m_{2}$.  The entropies
of these black holes are (resp.)
\bea
\mathscr{S}_{1} = \frac{1}{4}[\mathscr{A}_{1} + 2\theta_{2} X(\mathbb{C}_{1})]
\quad
\mbox{and}
\quad
\mathscr{S}_{2} = \frac{1}{4}[\mathscr{A}_{2} + 2\theta_{2} X(\mathbb{C}_{2})] \, ;
\eea
here we have defined the surface area $\mathscr{A}$ and correction term $X(\mathbb{C})$ via
\bea
\mathscr{A} = \oint_{\mathbb{C}}\tilde{\epsilon}
\quad
\mbox{and}
\quad
X(\mathbb{C}) = \oint_{\mathbb{C}}\tilde{\epsilon}\mathcal{R} \; .
\eea
Before the black holes merge, the total entropy is
\bea
\mathscr{S} = \mathscr{S}_{1} + \mathscr{S}_{2} = \frac{1}{4}[\mathscr{A}_{1} + \mathscr{A}_{2}
              + 2\theta_{2}(X(\mathbb{C}_{1}) + X(\mathbb{C}_{2}))] \; .
\eea
After the black holes merge, the total entropy of the resulting black hole is
\bea
\mathscr{S}^{\prime} = \frac{1}{4}[\mathscr{A}^{\prime} + 2\theta_{2} X(\mathbb{C}^{\prime})] \; .
\eea
The area theorem will hold if and only if $\mathscr{S}^{\prime}>\mathscr{S}$.  Thus we have
the following bound:
\bea
\theta_{2} < \frac{-(\mathscr{A}_{1} + \mathscr{A}_{2} - \mathscr{A}^{\prime})}{2[X(\mathbb{C})
            + X(\mathbb{C}_{2}) - X(\mathbb{C}^{\prime})]} \; .
\label{bound}
\eea
This bound for $\theta_{2}$ is general, and holds for spacetimes in all dimensions with generic values
of $\Lambda$.

Without knowing more about the details of the black holes that are merging, nothing further can be said
about the bound (\ref{bound}) because the topological structure of black holes is much richer in $D>4$
dimensions than in four dimensions.  When $\Lambda\geq0$ the topology of black holes in $D\geq5$
dimensions can be \emph{any} product manifold $\mathbb{R}^{2}\times\mathbb{C}^{D-2}$, with
$\mathbb{C}^{D-2}$ a space of \emph{positive Yamabe type}.  For example, in five dimensions the topology
of the event horizon has to be (a connected sum of) a three-sphere $\mathbb{C}^{3}\cong S^{3}$ or three-ring
$\mathbb{C}^{3}\cong S^{2}\times S^{1}$.  A complete discussion of black holes in higher dimensions is
presented in \cite{emprea}; topological properties are presented in \cite{caigal,galsco,galloway,hoy,racz,hhi}.

For concreteness, then, let us consider the merging of two non-rotating black holes in four dimensions, with
$\Lambda=0$.  First, the Gauss-Bonnet theorem says that
\bea
X(\mathbb{C}) = \oint_{\mathbb{C}}\tilde{\epsilon}\mathcal{R} = 4\pi\chi(\mathbb{C}) \, ,
\eea
with $\chi(\mathbb{C})$ the Euler characteristic of $\mathbb{C}$.  Then the Hawking topology theorem says
that the horizon cross sections can only be two-spheres so that $\mathbb{C}\cong S^{2}$, and then $\chi(S^{2})=2$.
It follows that
\bea
X(\mathbb{C}_{1}) = X(\mathbb{C}_{2}) = X(\mathbb{C}^{\prime}) = 8\pi \; .
\eea
Finally, the Birkhoff theorem says that the only static asymptotically flat solution to the field equations
is the Schwarzschild solution.  Since the surface area of a Schwarzschild black hole is related to its mass
$m$ via $\mathscr{A}=16\pi m^{2}$, the surface areas of the initial and final black-hole states are
\bea
\mathscr{A}_{1} = 16\pi m_{1}^{2} \, ,
\quad
\mathscr{A}_{2} = 16\pi m_{2}^{2} \, ,
\quad
\mbox{and}
\quad
\mathscr{A}^{\prime} = 16\pi(m_{1} + m_{2} - \alpha)^{2} \; .
\eea
In the above definition for $\mathscr{A}^{\prime}$ the parameter $\alpha\geq0$ has been added which corresponds
to any mass that may be carried away by gravitational radiation during merging.  Whence the bound on $\theta_{2}$:
\bea
\theta_{2} < 2m_{1}m_{2} - \alpha[2(m_{1} + m_{2}) - \alpha] \; .
\label{bound2}
\eea
Therefore, in four-dimensional asymptotically flat spacetimes, the second law will be violated if $\theta_{2}$
is greater than twice the product of the masses of two Schwarzschild black holes before merging minus a correction
due to gravitational radiation.  This is an important property because it shows that a non-zero Euler characteristic
of the manifold with boundary can have physical effects in four dimensions.



\end{document}